# Understanding Teams and Productivity in Information Retrieval Research (2000-2018): Academia, Industry, and Cross-Community Collaborations

Jiaqi Lei [+]

*Institute of Education, Tsinghua University, Beijing 100084, China*

*Department of Information Management, Peking University, Beijing 100871, China*

Liang Hu [+]

*Department of Information Management, Peking University, Beijing 100871, China*

Yi Bu *

*Department of Information Management, Peking University, Beijing 100871, China*

Jiqun Liu *

*School of Library and Information Studies, The University of Oklahoma, Norman, Oklahoma 73019, U.S.A.*

**Note: Jiaqi Lei and Liang Hu contributed to this paper equally.**

**Corresponding concerning this article should be addressed to Yi Bu ([buyi@pku.edu.cn](mailto:buyi@pku.edu.cn)) and Jiqun Liu ([jiqunliu@ou.edu](mailto:jiqunliu@ou.edu)).**

**Abstract**: Previous researches on the Information retrieval (IR) field have focused on summarizing progress and synthesizing knowledge and techniques from individual studies and data-driven experiments, the extent of contributions and collaborations between researchers from different communities (e.g., academia and industry) in advancing IR knowledge remains unclear. To address this gap, this study explores several characteristics of information retrieval research in four areas: productivity patterns and preferred venues, the relationship between citations and downloads, changes in research topics, and changes in patterns of scientific collaboration, by analyzing 53,471 papers published between 2000 and 2018 from the Association for Computing Machinery (ACM) Digital Library dataset. Through the analysis and interpretation on empirical datasets, we find that academic research, industry research, and collaborative research between academia and industry focused on different topics. Among the collaboration models, Academia-Industry Collaboration is more oriented towards large teamwork. Collaborative networks between researchers in academia and industry suggest that the field of information retrieval has become richer over time in terms of themes, foci, and sub-themes, becoming a more diverse field of study.

# INTRODUCTION

Information retrieval (IR) research seeks to characterize, support, and improve the process of retrieving relevant information and documents that satisfy users' information needs (Kobayashi & Takeda, 2000). It is an interdisciplinary research field that emphasizes the value of evaluation and brings together knowledge and methods from computer science, library and information studies, human-computer interaction, and other related areas. IR attracts attention and research efforts from both academia and industry, fostering interdisciplinary collaborations. In the past two decades, as search systems become increasingly ubiquitous in different modalities of human-information interactions (such as desktop search, mobile search, conversational/spoken information seeking and chat search), both researchers and system developers from academia and industry have made significant contributions to IR algorithms, interactive search systems, user models, as well as evaluation techniques. Furthermore, the area has witnessed a resurgence of interest in natural language processing and deep learning. Artificial Intelligence (AI) techniques mark a series of unique contributions from industry researchers to the area for developing and evaluating modern intelligent search systems (Culpepper et al., 2018; Li & Lu, 2016; Yates et al., 2021).

While previous surveys and workshops have focused on summarizing progress and synthesizing knowledge and techniques from individual studies and data-driven experiments, the extent of contributions and collaborations between researchers from different communities (e.g., academia and industry) in advancing IR knowledge remains unclear. To address this gap, this study aims to answer the following four research questions. This paper begins by focusing on a straightforward question of whether scientists from academia and industry have preferences and differences in their choice of academic conferences for publishing their papers. Thus, the first research question is formulated as follows:

RQ1: What are the patterns of productivity and preferred venues characterize information retrieval studies by Academia, Industry, and Academia-Industry Collaboration?

The number of full-text paper downloads as an important element of alternative metrics of traditional citation-based measures has now received attention from scholars worldwide. Garfield (1996) proposed the use of web downloads instead of citations of scientific publications to resolve the problem that there are temporal lags in the evaluation of scientific publications using citation analysis. Due to the timeliness of downloads and their complementary effect on citations, previous studies have explored the relationship between download and citation counts (Hu et al., 2021). This paper specifically investigates the relationship between citation and download counts between academia and industry in the field of information retrieval, and introduces download conversion rates (the ratio of citations to downloads of a paper) and gender-related factors for further investigation. Consequently, the second research question is as follows:

RQ2: What is the relationship between citation and downloads counts in Academia, Industry, and Academia-Industry Collaboration?

Given the distinct working culture and various orientations of academia and industry, researchers from academia may tend to lean towards theoretical research while those from industry may focus more on practical applications and systems. That being said, we may expect to observe significant differences in the topics covered by these researchers. When researchers from academia and industry collaborate, one of the issues this paper will explore is the changing focus of research attention and emerging research topics. Hence, the third research question is proposed:

RQ3: How do the research topics change over time in the three types of papers: Academia (all co-authors are from academia), Industry (all co-authors are from industry), and Academia-Industry Collaborations (some co-authors are from academic while others industry)?

As one of the focuses of this paper is on the collaborative outputs from academia and industry, we are eager to know what types of papers are more likely to involve large-team collaborations and what the characteristics of author team sizes in an Academia-Industry collaborations are. Through statistical analysis on author teams, this paper aims to explore these and provide reasonable explanations. Additionally, understanding the changing trends in collaboration between academia and industry is an important aspect of this research. Thus, the final research question is as follows:

RQ4: What is the preferred size of collaboration teams in the three categories of papers, and how does collaboration between authors from academia and industry evolve over time?

Findings from the current study may offer a new perspective for analyzing the advance and emerging trends in IR research and helps clarify the cross-community collaborations and scientific contributions of academia and industry. In what follows, we first review the relevant papers, introduce the data sources and the data process of this paper, and finally answer each of the four research questions in the results section. At the end, we will summarize and discuss the results.

## RELATED WORK

*Academia-industry collaborations*

Previous studies have noted discrepancy between researches from academia and industry. Academia typically tends to focuses on basic research and scientific exploration while industry is driven by commercial purposes (Ahmed et al., 2023). Academia places more emphasis on scientific breakthroughs and innovation in knowledge, while industry links research and development activities to market needs and commercial interests (Spicer et al., 2022).

In recent years, collaborative behavior between academia and industry has become increasingly common (Wuchty et al., 2007; Zhang et al., 2018). However, cultural differences between academia and industry may make this collaborative system difficult and challenging. Jasny et al. (2017) explore such collaborative systems where incomplete communication and sharing of technology, data, or materials interfere with future research, and advocate for leadership, and support from funding agencies, journals, and other stakeholders. Marijan and Gotlieb (2021) address the challenges in establishing effective scientific collaboration between academia and industry, the Certus model was proposed to facilitate participatory knowledge creation when solving problems. Furthermore, recent studies indicate that the gap in research collaboration between academia and industry is progressively narrowing (Etzkowitz & Leydesdorff, 2000; Rhoten & Powell, 2007). This trend is driven by a growing recognition of the value of interdisciplinary approaches, increased support from funding agencies and research institutions, and advancements in technology that enable seamless communication and knowledge sharing.

Despite the difficulties and challenges, the academic-industrial collaboration model has great merit. Collaboration between academia and industry can "translate" scientific discoveries into industrial impact, commercializing researches that would otherwise go undiscovered. In addition, collaboration between academia and industry can promote knowledge sharing and technology transfer, and improve the application of researches (Noyons et al., 1994; Perkmann & Walsh, 2009). The industry community gain new business opportunities and competitive advantages from the research results of academia, and academia community obtain more research resources and financial supports (Owen-Smith, 2003; Van Looy et al., 2006). In short, cooperation between academia and industry can maximize the value of researches and promote the development of IR researches.

*Emerging topics and collaborations in Information Retrieval research*

Previous research in the field of Information Retrieval (IR) has encompassed a variety of approaches, such as user studies, simulation-based experiments, and naturalist studies, which have been employed to address diverse unresolved challenges and emerging problems both within and beyond technical, system-oriented aspects of IR. Keyvan and Huang (2023) conducted a survey focusing on techniques, tools, and methods used to comprehend ambiguous queries in Conversational Search Systems (CSS) deployed in everyday-life and workplace settings, such as chatbots, Apple's Siri, Amazon Alexa, and Google Assistant. Ambiguous query clarification and search result re-ranking, among other open questions, have been extensively explored and discussed in publications from academia, industry, and collaborative projects involving both sides (e.g., (J. Gao et al., 2020; Thomas et al., 2021; Zamani et al., 2020).

Recently, the emergence of algorithmic fairness, accountability, transparency, and ethics (FATE) as a notable research topic has attracted attention from both academic and industry scholars in the IR community. This line of research has resulted in a series of publications, industry sessions, collaborative workshops, tutorials, and funding projects (Castillo, 2019; Ekstrand et al., 2019; R. Gao & Shah, 2021). The FATE-IR research has brought together a diverse group of researchers and practitioners who contribute to both the conceptualization and technical aspects of responsible IR research agenda (Olteanu et al., 2019).

Currently, there are few articles examining the collaboration between academia and industry in the field of IR. (Zaharia & Kaburakis, 2016) explore trends in collaboration barriers among various research involvement levels of U.S. sport firms with sport management academia. (Ahmed et al., 2023) examine the current state of research in AI from industry and academia working together to tip the scales in favor of industry. A previous research-in-progress paper of the current one examines the features and differences regarding productivity, authorship, and impact of the three types of studies and also pay special attention to the research problems and topics that attract and foster academia-industry collaborations in the recent two decades of IR studies (Lei et al., 2023). Built on the preliminary analysis of the collaboration between academia and industry in IR in terms of productivity, authorship, impact, and topic distribution, the current paper will leverage an extended dataset, aiming to answer the four research questions (RQs) proposed above.

## DATA PREPARATION

The empirical data employed in our analysis mainly comes from Association for Computing Machinery (ACM) Digital Library[1], which is a comprehensive repository of articles in the field of computer science and related areas. The original dataset we utilized comprised a total of 295,561 articles published in ACM from 1951 to 2018. The dataset includes information such as the publication date, title, abstract, keywords, author and author institution IDs and names, citation count, and download count for each paper. To filter out articles in the field of information retrieval, we identified a set of 255 keywords (detailed results are provided in Table A1) from the whole keywords set. These keywords were used for matching in the article keyword field in the original dataset, which ultimately matched 53,471 articles published between 2000 and 2018, referring to this dataset as the ACM dataset.

The Research Organization Registry (ROR)[2] is an inclusive and community-driven global registry that maintains open persistent identifiers for research organizations. The ROR dataset provides the categories of us to obtain the types of author institutions, which allows us to classify authors into three categories: academia, industry, and others (e.g., facilities, health, governments, etc.). This classification process was facilitated by extracting 137,843 (author, institution) data pairs from the ACM dataset and matching them with the corresponding entries in ROR dataset. Our matching efforts successfully identified 125,668 data pairs, accounting for 91.17% of the total data. However, there remained 8.83% of the data pairs that did not find a direct match. To ensure that no significant institutions were overlooked, we took a proactive measure. Specifically, we manually supplemented the type labels for institutions that appeared 10 times or more in the dataset. This supplementation involved conducting ROR searches to assign appropriate type labels. As a result, only 3.69% of the author institutions did not have a corresponding type match.

---

[1] https://dl.acm.org
[2] https://ror.org

We conducted further classification of scientific publications into four distinct types: publications authored/co-authored exclusively by individuals from academia (Academia), publications authored/co-authored exclusively by individuals from industry (Industry), publications co-authored by individuals from both academia and industry (Academia-Industry Collaboration), and others. Our analysis revealed that within the ACM dataset, there were 37,034 papers classified as Academia, 4,941 papers classified as Industry, 1,986 papers classified as Academia-Industry Collaboration, and 2,604 papers classified as other types. These numbers indicate that the majority of papers in the field of information retrieval are authored/co-authored by individuals within the academic community, followed by the number of papers published collaboratively between academia and industry, and the type with the fewest number of papers published entirely by researchers from the industrial community. This paper primarily focuses on the first three types, which collectively comprise a total of 46,565 papers. The specific data processing steps are illustrated in Figure 1.

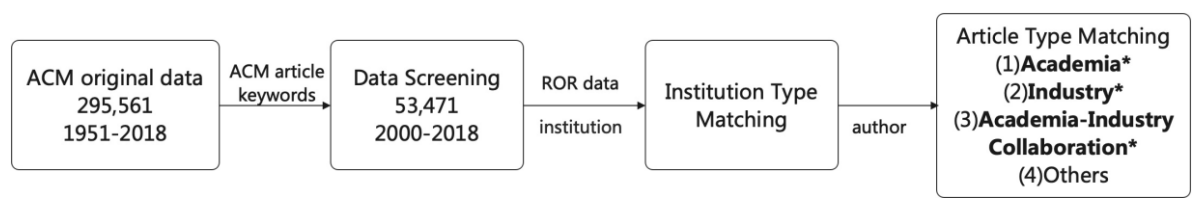

Figure 1. Flow chart of data processing. * indicates the focus of this current paper

Due to the incomplete nature of the ACM dataset, some IR reviewed conferences may not be included in our dataset. Papers from some of the refereed conferences may represent high-quality research and recent advances in the field of information retrieval. The exclusion of these papers may result in a dataset that lacks the necessary standardization and consistency in evaluating and comparing information retrieval technologies. However, the ACM digital libary includes conference papers from as comprehensive a range of IR fields as possible, and most papers in the IR field are presented at conferences. Even if all papers published in refereed conferences are excluded, conference papers included in the dataset may come from more than one conference, and this diversity helps researchers gain a broad understanding of different conferences, different research communities, and different research topics.

This paper categorizes authors according to the type of institution they belong to. The names of the authors and their institutions at the time of publishing the corresponding paper are given in the original dataset. Therefore, while the authors' institutional affiliations and their categories may have changed over their academic careers, they will change accordingly in our dataset. To ensure the accuracy of the data, we also extracted a portion of the data for manual inspection.

## ANALYSIS AND RESULTS

This section focuses on presenting the findings obtained from the analysis of the pre-processed ACM publication dataset discussed earlier. The research questions are addressed across four main areas, namely productivity patterns and preferred venues for three types of articles, the relationship between the number of citations and downloads of the three types of papers and the derived questions, changes in dissertation research topics over time, changes in the number of partners and variations in the number of partnerships and changes in partnerships.

In the *productivity patterns and preferred venues* part, we present visual representations of the top ten published conferences for the three types of articles and compare their similarities and differences. In the *citations and downloads* part, We firstly explore the correlation between citations and downloads for the three types of articles using heat maps of the correlation coefficient matrix and an indicator named conversion rate. Furthermore, a negative binomial regression model is applied to explore the impact of gender composition on the number of citations and downloads of IR papers. In the *research topic analysis* part, a pre-trained BERT is employed to extract keywords and get the five most important words for each year and each paper type to investigate the potential changes in research topics over time. In the *scientific collaborations* part, we firstly do some team size-wise explorations. Then we perform co-authorship network analysis by constructing the co-authorship networks and visualizing the networks in a temporal manner with Gephi. Within each subsection, we provide an introduction to the methods employed and the corresponding results for each research question.

*Patterns of productivity and preferred venues*

Figure 2 presents visual representations of the top ten published conferences within each of the three categories: Academia, Industry, and Academia-Industry Collaboration. The pie charts provide a clear overview of the distribution of publications among these conferences. The top left figure represents Academia, the top right figure represents Industry, and the bottom figure represents the Academia-Industry Collaboration. Conferences are shown in the legend from top to bottom in descending order of representation. In the three categories, CHI (Conference on Human Factors in Computing Systems), WWW (International World Wide Web Conference) and CIKM (Conference on Information and Knowledge Management) hold the top three positions. However, their rankings vary within each category. Interestingly, WWW emerges as the top conference in both the Industry and Academia-Industry Collaboration categories. Furthermore, the top five conference rankings in these two categories remain consistent. This observation suggests that when a co-author from industry is involved, the paper's content is more likely to have an industrial focus, leading to a preference for conferences aligned with purely industrial papers.

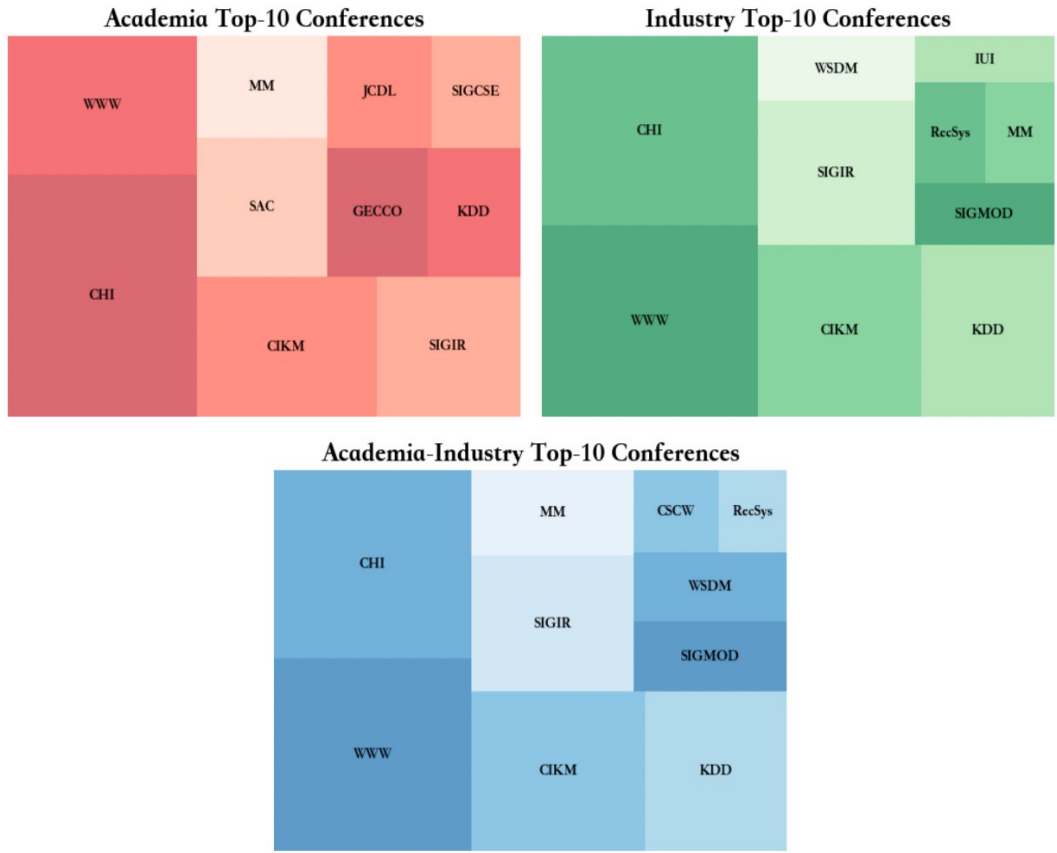

**Figure 2. The top-10 published conferences in the three categories**

Notably, two conferences, RecSys (The ACM Conference Series on Recommender Systems) and CSCW (Conference on Computer supported cooperative work), only appear in the top ten list of conferences for the Academia-Industry Collaboration category. This finding indicates that these conferences are more prevalent in collaborative research efforts between academia and industry. Overall, these insights shed light on the conference preferences and content orientations within each category, highlighting the influence of industry collaboration on the publication choices and directions of research papers.

*Citations and downloads*

Correlation between citations and downloads

With the development of the Internet, the electronicization of academic papers is becoming more and more popular, and almost all journal papers are able to be accessed through online databases. Before a paper is cited, there will be browsing, downloading, reading and other use behaviors, of which downloading behavior is more easily recorded and targeted. Therefore, downloads have gradually become one of the mainstream and important alternative measures (Schloegl & Gorraiz, 2011), Supplementing the analysis of downloads and citations enables a more comprehensive assessment of the impact of the focal paper, and also helps to understand the dynamics of the paper's dissemination and acceptance process, thus facilitating more effective academic communication and knowledge dissemination. To initially explore the relationship between the number of citations and downloads, we calculated the Pearson's correlation coefficients between the three types of data and visualized them in heat maps, as shown in Figure 3In the figure, citation_count, downloads_cu, downloads_12month, and downloads_6week represent the total number of citations, the total number of downloads, the number of downloads in the last year, and the number of downloads in the last six weeks (during the time frame from which the dataset is acquired). The results of the Academia, Industry, and Academia-Industry Collaboration are shown from the top to the bottom panels of Figure 3 It is evident that Academia and Academia-Industry Collaboration exhibit similar patterns and regularities while Industry shows distinct differences. Additionally, the figure reveals a strong correlation between the number of citations and the cumulative number of downloads for all three types of papers, with correlation coefficients exceeding 0.60. This suggests that the number of citations and cumulative downloads can serve as predictors for each other. In particular, the correlation coefficients between downloads within six weeks and downloads within one year are particularly high for all the three types of papers (greater than 0.7). In the case of Industry, this coefficient even reaches 0.94. However, when examining the correlation between the total number of downloads and the number of downloads within six weeks or within one year, the coefficients are relatively small. This indicates that the significant variation in downloads over time is not adequately captured by the total download count. Moreover, the correlation between citations and downloads strengthens when a team include scientists from academia. Conversely, in the case of Industry, the correlation between citations and downloads within six weeks surpasses the correlation between downloads within one year.

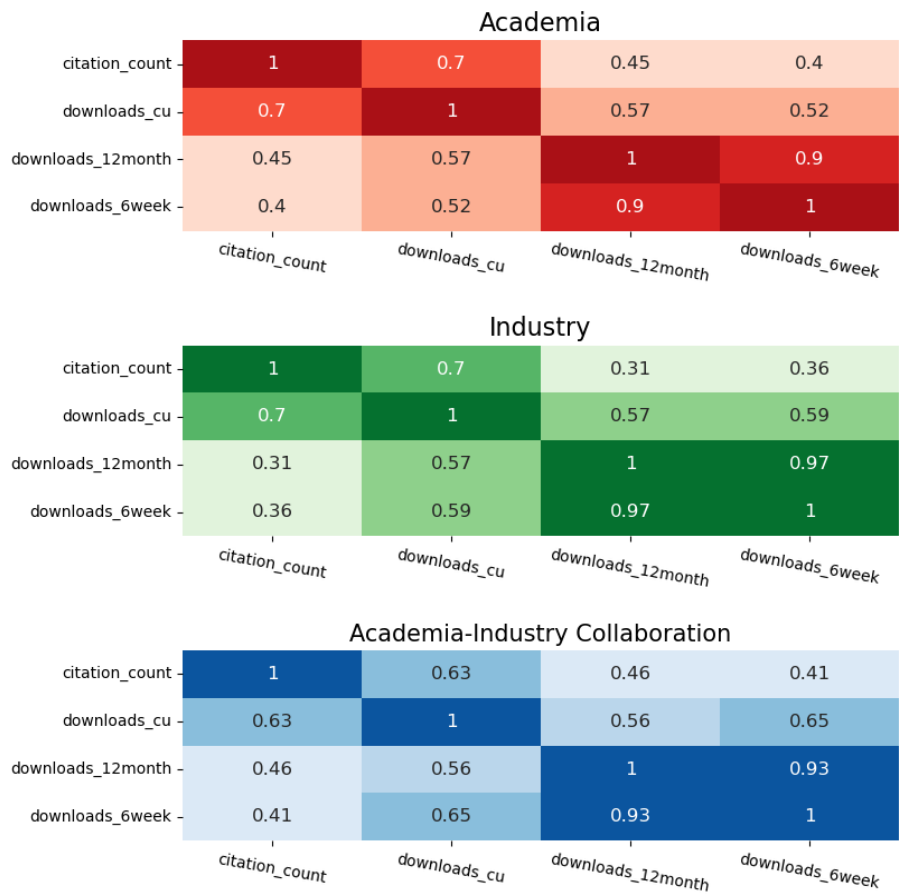

**Figure 3. Heat map of the correlation coefficient matrix for the three types of articles.**

Conversion Rates

To further explore the relationship between downloads and citations, we introduced a simple metric, namely conversion rate, which is calculated as:

$$CR_i = citations_i/downloads_i \quad (1)$$

In Equation 1, $CR_i$ stands for conversion rate and $i$ refers to a scientific publication. $citations_i$ and $downloads_i$ indicate the number of citations and downloads of publication $i$. This metric is used to measure how many cumulative downloads of each article will be converted into real citations. To present the findings more effectively, raw data was grouped into 15 bins (Figure 4). Figure 4 clearly demonstrates that scientific publications originating from Academia-Industry Collaboration consistently exhibit higher conversion rates compared to articles purely from Academia and purely from Industry, regardless of the number of downloads. This suggests that, for an equivalent number of downloads, articles by Academia-Industry Collaborations are more likely to receive citations. These results provide invaluable insights into the impact of collaborative efforts between academia and industry on citation rates. A greater conversion rate observed in the Academia-Industry Collaboration category may imply the added value and influence of this collaborative research approach.

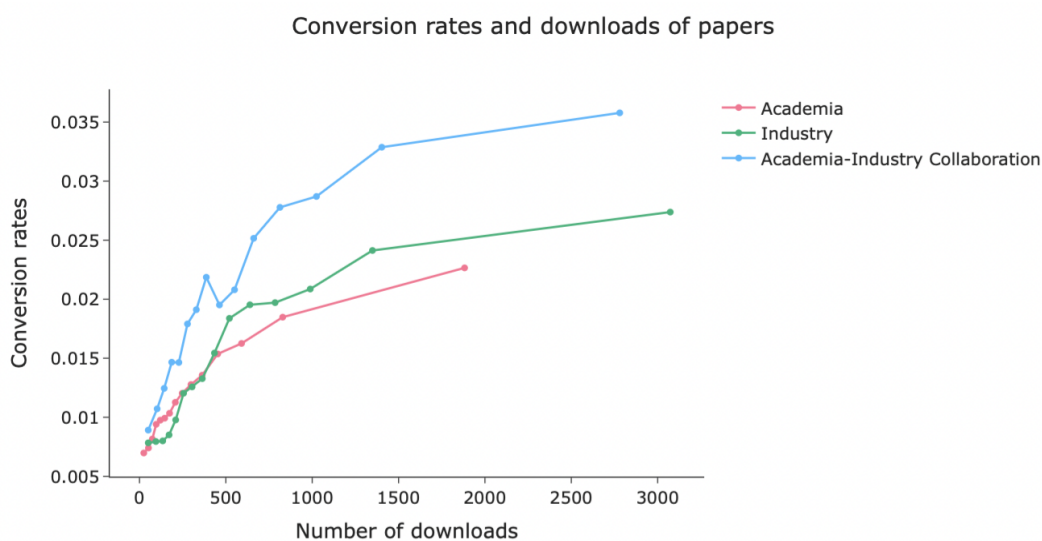

**Figure 4. Conversion rate for the three types of articles.**

Gender, citations, and downloads

Related studies indicate that gender may impact the visibility and impact of scientific outputs (Johnson, 1987; Sotudeh & Khoshian, 2014). This research aims to reveal gender biases in the scientific environment of the information retrieval field by examining how gender composition affects the citations and download of papers. To explore the impact of gender composition on the number of citations and downloads of IR papers, we firstly utilize a Python package named gender-guesser to identity the gender of the authors[3]; this package has been widely adopted in gender identification tasks (Adler et al., 2020; Shang et al., 2022; Zeina et al., 2020). First, we plotted the gender distribution of authors across different types of papers and observed that in all types, the number of male authors significantly exceeds that of female authors as shown in **Figure 5**. Then a negative binomial regression model is applied:

$$ln\ citation\ number_{i,t}\ (ln\ download\ number_{i,t}) = \beta_o + \beta_1 \cdot gender\ ratio_i + \beta_2 \cdot paper\ type_i + \beta_3 \cdot team\ size_i + \delta_t + \theta_i + \varepsilon \quad (2)$$

The model contains dependent variables, i.e., logarithmic citation counts or cumulative download number of paper *i* at time *t*, and an independent variable, gender ratio. We define the gender ratio of paper *i* as the ratio of female author numbers to male author numbers. It is worth noting that papers with no male author are not included as regression data when gender ratio is used. Moreover, the type and the team size of paper *i* are included as control variables, and the year when paper *i* published, $\delta_t$, and the topic of paper *i*, $\theta_i$, are included as fixed effects, to control the time- and topic-invariant factors. There are four types of publications, academia, industry, collaboration and others. We define three binary variables to represent industry, collaboration and others, respectively. For example, if the type of a publication $i$ is industry, then the binary variables industry$_i$ = 1, collaboration$_i$ = 0, and others$_i$ = 0. And if the type is academia, then the values of all the three binary variables are 0. As for paper topic, we utilize the LDA topic model (Blei et al., 2003) and divide the papers into nine topics according to perplexity index (Figure A1). As a robustness test, we change the gender ratio into female ratio, namely the ratio of the number of female authors to the total number of authors in a team, in the regression model.

---
[3] https://pypi.org/project/gender-guesser/

$$\ln citation\ number_{i,t}\ (\ln download\ number_{i,t})$$
$$= \beta_0 + \beta_1 \cdot gender\ ratio_i + \beta_2 \cdot paper\ type_i + \beta_3 \cdot team\ size_i + \delta_t + \theta_i + \varepsilon$$

The descriptive statistics of the variables is shown in Table 1. We also calculate the correlation coefficient matrix of the variables in Table 2. From the correlation coefficient matrix, we find that the correlation between most variables is not strong, except gender ratio and female ratio.

As shown in Models 1 and 2 in Table 3, both the numbers of citations and downloads are negatively correlated with gender ratio ($p<0.1$ and $p<0.01$, respectively). Specifically, in Model 1, the number of citations becomes 97.2% ($e^{-0.028}$) times the original for every 1 increase in gender ratio. The greater the absolute value of the negative coefficient, the stronger the negative correlation. When the independent variable is changed into female ratio (instead of gender ratio) in Models 3 and 4, the dependent variables are still negatively correlated, which may suggest that female scientists are relatively disadvantaged in scientific research regardless of Academia, Industry, or Academia-Industry collaboration.

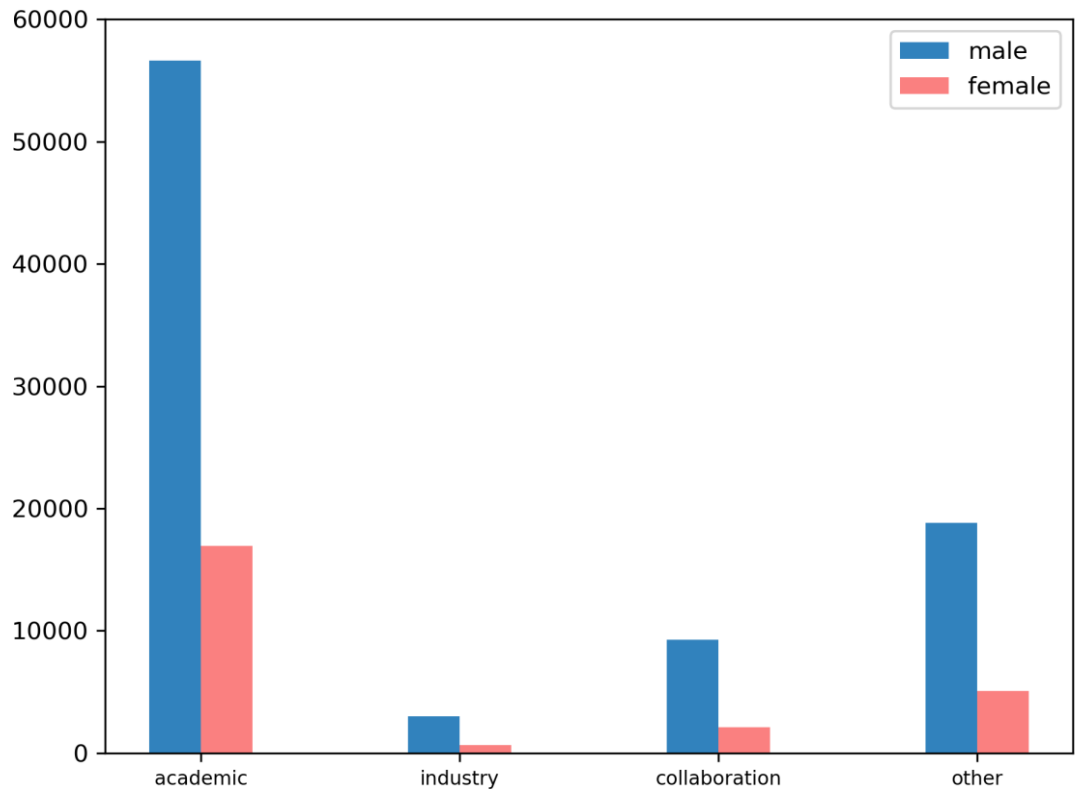

**Figure 5. Gender distribution of authors across different types of papers.**

**Table 1. Descriptive statistics of the variables.**

| VarName | Obs | Mean | SD | Median | Min | Q1 | Q3 | Max |
|---|---|---|---|---|---|---|---|---|
| citations | 53470 | 8.026 | 25.897 | 2.000 | 0.000 | 0.000 | 6.000 | 1399.000 |
| downloads | 52948 | 411.578 | 703.782 | 226.000 | 0.000 | 111.000 | 470.000 | 49889.000 |
| gender_ratio | 41057 | 0.298 | 0.563 | 0.000 | 0.000 | 0.000 | 0.500 | 8.000 |
| female_ratio | 53470 | 0.148 | 0.250 | 0.000 | 0.000 | 0.000 | 0.250 | 1.000 |
| industry | 53470 | 0.037 | 0.189 | 0.000 | 0.000 | 0.000 | 0.000 | 1.000 |
| collaboration | 53470 | 0.092 | 0.290 | 0.000 | 0.000 | 0.000 | 0.000 | 1.000 |
| other | 53470 | 0.178 | 0.382 | 0.000 | 0.000 | 0.000 | 0.000 | 1.000 |
| team_size | 53470 | 3.266 | 1.680 | 3.000 | 1.000 | 2.000 | 4.000 | 76.000 |

**Table 2. Pearson's correlation matrix.**

| | citations | downloads | gender_ratio | female_ratio | industry | collaboration | other | team_size |
|---|---|---|---|---|---|---|---|---|
| citations | 1 | | | | | | | |
| downloads | 0.706*** | 1 | | | | | | |
| gender_ratio | -0.011** | -0.028*** | 1 | | | | | |
| female_ratio | -0.012** | -0.030*** | 0.991*** | 1 | | | | |
| industry | 0.041*** | 0.085*** | -0.041*** | -0.040*** | 1 | | | |
| collaboration | 0.164*** | 0.156*** | -0.002 | -0.010* | -0.065*** | 1 | | |
| other | -0.001 | -0.015*** | 0.045*** | 0.047*** | -0.096*** | -0.158*** | 1 | |
| team_size | 0.089*** | 0.035*** | 0.240*** | 0.216*** | -0.069*** | 0.151*** | 0.134*** | 1 |

\*\*\* p<0.01, \*\* p<0.05, \* p<0.1

Table 3. Regression results.

|  | Model 1 citations | Model 2 downloads | Model 3 citations | Model 4 downloads |
|---|---|---|---|---|
| gender_ratio | -0.028* | -0.033*** |  |  |
|  | (-1.892) | (-4.153) |  |  |
| female_ratio |  |  | -0.185*** | -0.107*** |
|  |  |  | (-5.890) | (-6.765) |
| industry | 0.429*** | 0.406*** | 0.416*** | 0.395*** |
|  | (10.605) | (17.979) | (11.332) | (19.637) |
| collaboration | 0.755*** | 0.473*** | 0.756*** | 0.450*** |
|  | (28.212) | (31.581) | (30.861) | (33.396) |
| other | -0.030 | -0.015 | -0.039** | -0.015 |
|  | (-1.436) | (-1.298) | (-2.041) | (-1.460) |
| team_size | 0.145*** | 0.073*** | 0.169*** | 0.078*** |
|  | (26.910) | (27.164) | (34.169) | (32.409) |
| time fixed effects | YES | YES | YES | YES |
| topic fixed effects | YES | YES | YES | YES |
| constant | 2.599*** | 6.454*** | 2.510*** | 6.432*** |
|  | (27.727) | (120.398) | (29.941) | (136.621) |
| ln alpha | 0.782*** | -0.303*** | 0.829*** | -0.293*** |
|  | (96.129) | (-47.284) | (114.182) | (-52.125) |
| observations | 41032 | 40643 | 53433 | 52911 |
| pseudo $R^2$ | 0.0670 | 0.0312 | 0.0690 | 0.0310 |
| log likelihood | -108393.26 | -277449.17 | -136174.84 | -359942.91 |

*** p<0.01, ** p<0.05, * p<0.1. t values are shown in parentheses.

*Research topic analysis*

To investigate the potential changes in research topics among the three types of articles over time, particularly for papers resulting from collaborations between academia and industry, we employ a large-scale pre-trained language model called DistilBERT for keyword extraction (Sanh et al., 2020). We first divide articles by their types and publication years and combine the title and abstract of an article into one document. We then utilize the sentence transformer model (Sanh et al., 2020) to encode the document where we select the keyword length for candidate keyword extraction as two. After that, the cosine similarity between candidate keywords and documents is calculated, based on which we select the top five keywords with the greatest similarity. Finally, we get the five most important words for each year and each paper type, as shown in Table A2. Overall, we found that academic research tends to involve topics of diverse types, whereas industry research usually focuses on specific platforms, data formats, and commercial applications. In addition, compared to academic research that main focuses on digital libraries and algorithmic improvement, industry research has studied a series of practical challenges related to multimedia content processing (e.g., video, music), Web search, and social media platforms. With respect to the collaborations between two communities, the topics involve both algorithmic studies and tool- or dataset-specific experiments, which are likely to bring together the scientific research strength of academia and industry's advantages in tools, platforms, and datasets generated by large user pools. In recent years, academia-industry collaborative research also pays increasing attention to human-centered topics and methods, such as online and social media bullying, chatbot developers, and crowdsourcing analytics. This phenomenon is aligned with recent growing trends in human-centered computing research that emphasizes not only algorithmic and system-oriented effectiveness, but also individual users' experiences, fairness and ethics, as well as broader cultural and societal impacts.

We then pay particular interests in the semantic similarity (measured by the cosine distance between two vectors representing the publications) among the three types of publications, demonstrated in Figure 6. While no obvious increasing or decreasing trends are observed in any of the categories, the similarity values between the categories range predominantly between 0.5 and 0.8. These findings suggest that research conducted within academia and industry may either align or diverge in preferences from year to year. However, collaborative papers between academia and industry contribute to "shifting" these dynamics, potentially resulting in either increased or decreased similarities between the two.

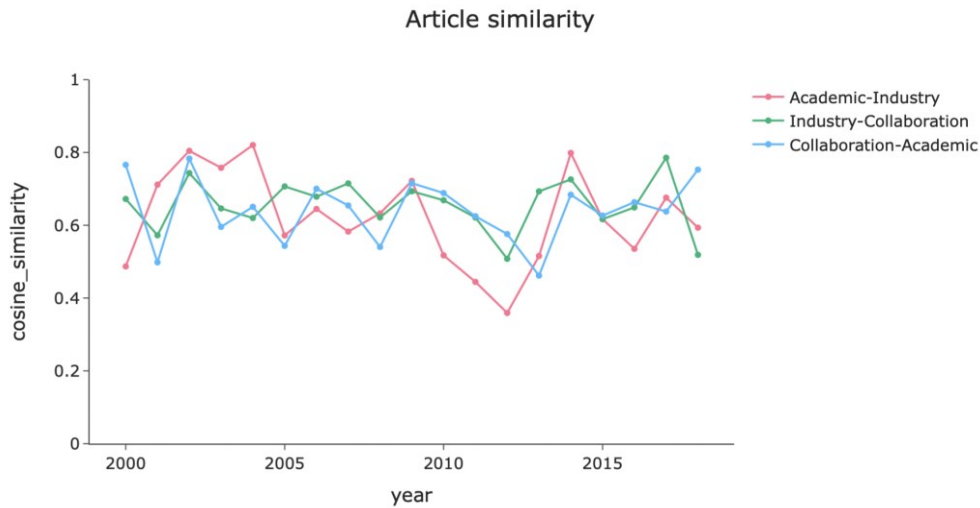

**Figure 6. Variation of cosine similarity with year for three types of articles.** "Academic-industry" indicates the similarity between publications by authors purely from academia and publications by authors purely from industry. "Industry-collaboration" indicates the similarity between publications by authors purely from industry and publications co-authored by scientists from academia and industry. "Collaboration-academic" refers to the similarity between publications co-authored by scientists from academia and industry and publications by authors purely from academia.

*Scientific Collaborations*

Figure 7 shows the percentage of the number of co-authors for the three types of papers in the form of a line graph. The proportion of single-author articles published by industry is 27.29%, significantly greater than that by academia (12.12%). This observation implies that researchers from industry have a stronger inclination towards conducting independent research. On the other hand, when the number of co-authors exceeds four, the percentage of Academia-Industry Collaboration consistently surpasses that of both Industry and Academia, which suggests that when academia-industry collaborations mostly occur in large teams.

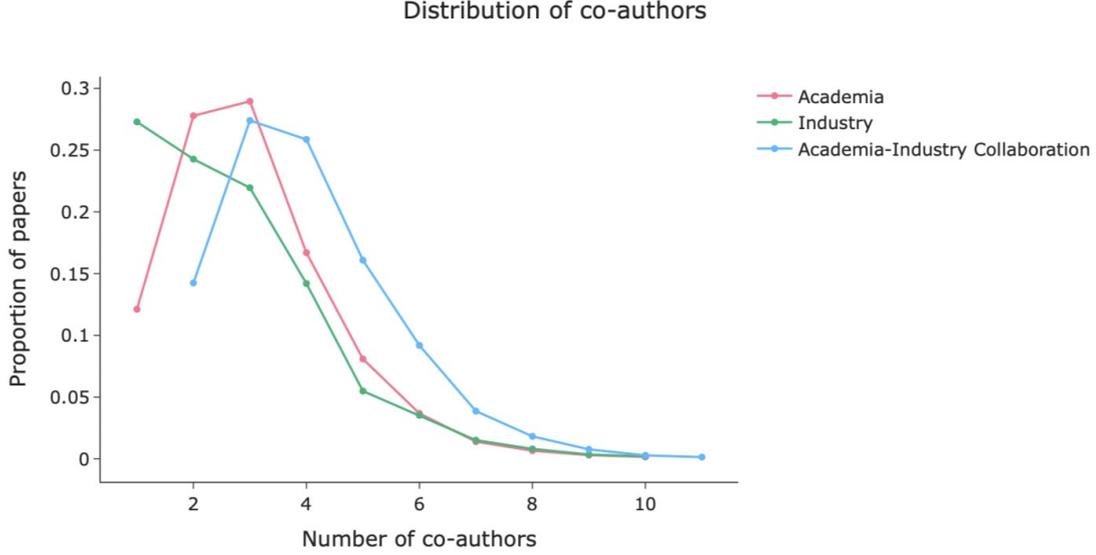

**Figure 7. Distribution of the number of co-authors for each type of publications. Since the Academia-Industry Collaboration is defined as having at least one author from academia and one author from industry, the blue curve representing Academia-Industry Collaboration starts with a number of co-authors of two.**

Besides team size, we further explore how the share of researchers from industry in each category has changed over time. This motivates us to introduce a new indicator, namely relative information entropy (RH), which is calculated by Equation 3.

$$RH = AH/TH \quad (3)$$

where $AH = -\sum_{j=1}^{s} \frac{N_j}{N} \times log_2 \frac{N_j}{N}$ is the actual information entropy (the actual entropy value of the distribution of authors from industry in the category) and $TH = -\sum_{j=1}^{s} \frac{N_j I}{N} \times log_2 \frac{N_j I}{N}$ refers to the theoretical information entropy (i.e., an ideal state [baseline]) where authors from industry are proportionally distributed in the category. $N$ represents the total number of authors in a team, $I$ represents the total number of authors from industry, $N_1$, $N_1$... and $N_s$ represents the number of authors in each category (in total s categories), $I_1$, $I_2$... and $I_s$ represents the number of authors from industry in each category. The values of AH, TH, and RH are shown in Table 4. We observe that the value of RH is decreasing, indicating that more equality regarding academia/industry in IR teams.

Table 4. RH calculation results for each cohort.

| Time cohort | TH | AH | RH |
| --- | --- | --- | --- |
| 2000-2004 | 6.149 | 3.750 | 0.610 |
| 2005-2009 | 5.364 | 3.692 | 0.688 |
| 2010-2014 | 4.800 | 2.568 | 0.535 |
| 2015-2018 | 5.553 | 3.086 | 0.556 |

Note that the detected trend of increasing and then decreasing RH values could be attributed to the initial screening process, where authors from industry with low publication numbers were excluded, leading to a bias towards industry in the network. Therefore, it is important to consider the limitations of the screening process and not to make generalizations about the decrease in industry participation or lack of significant influence solely based on these results. The specific screening and clustering results are shown in Table A4. In order to make the visualization more meaningful and readable, we filter the nodes and cluster the final network.

Coauthorship networks are usually regarded as a proximity of scientific collaborations (Milojević, 2010). Besides team size-wise explorations, we then perform coauthorship network analysis by dividing the dataset into four time cohorts and selecting the data of papers published in each cohort to construct the co-authorship network. We visualize coauthorship networks in a temporal manner with Gephi (from Figure 8 to Figure 11) where nodes represent authors and their size indicates the total number of publications by that author in each time period (Bastian et al., 2009). Edges represent co-authorship relations, and their width is proportional to the number of co-authored publications by both authors in each cohort. In each figure, the top panel includes the results after calculating modularity where the Community detection function utilizes a modularity-based community discovery algorithm along with the Louvain algorithm (Blondel et al., 2008). Each color in the top panel represents a specific category, referred to as the focus category, while grey is used to denote the other classes or categories. The lower panels in each figure shows the coauthorship network between academia and industry in the same position, labeling researchers in academia as red and researchers in industry as green.

Overall, we found that more dense edges/connections, sub-clusters, and widely distributed sub-centers emerge as the visualized co-authorship network evolve over time, indicating that IR has been significantly enriched with increasingly diverse topics, focuses, and sub-areas. In addition, we observe that, as the research field moves forward, the connections and overlaps between academia and industry authors increase, suggesting that the shared interests and expertise between authors from the two communities grow over time.

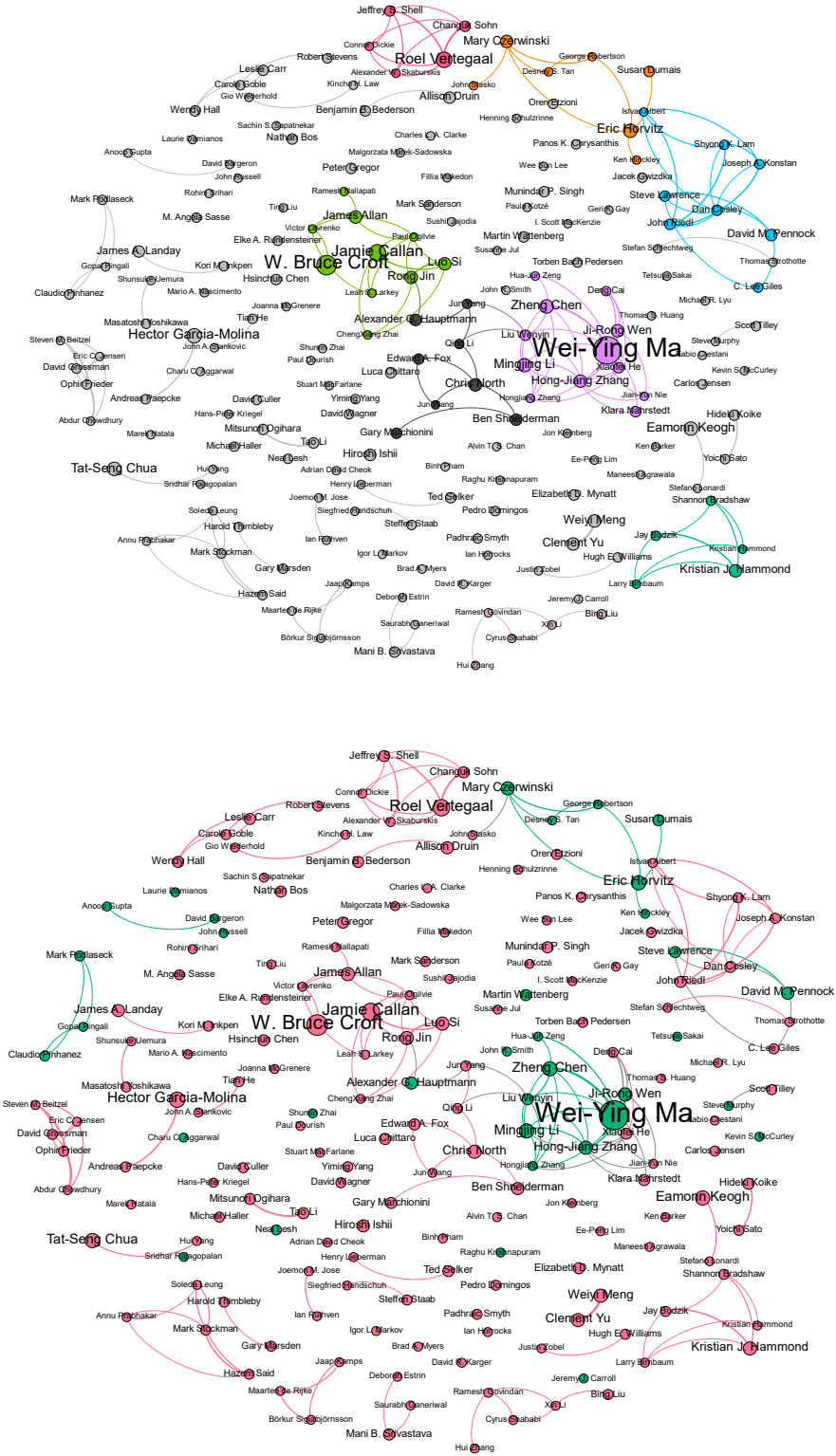

**Figure 8. Co-authorship network 2000-2004 (Top: node color by modularity; bottom: node color by authors' categories).**

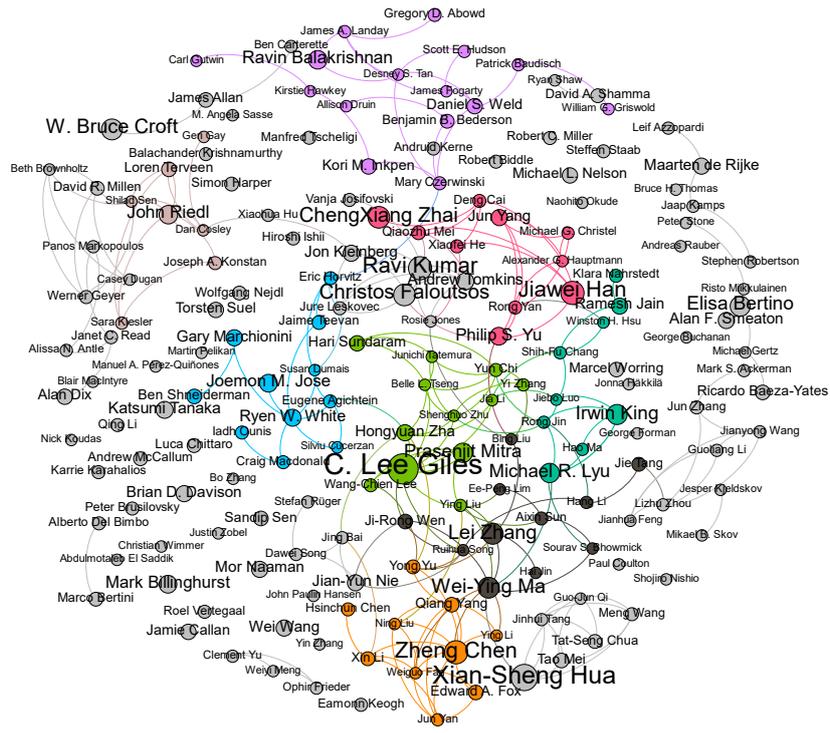
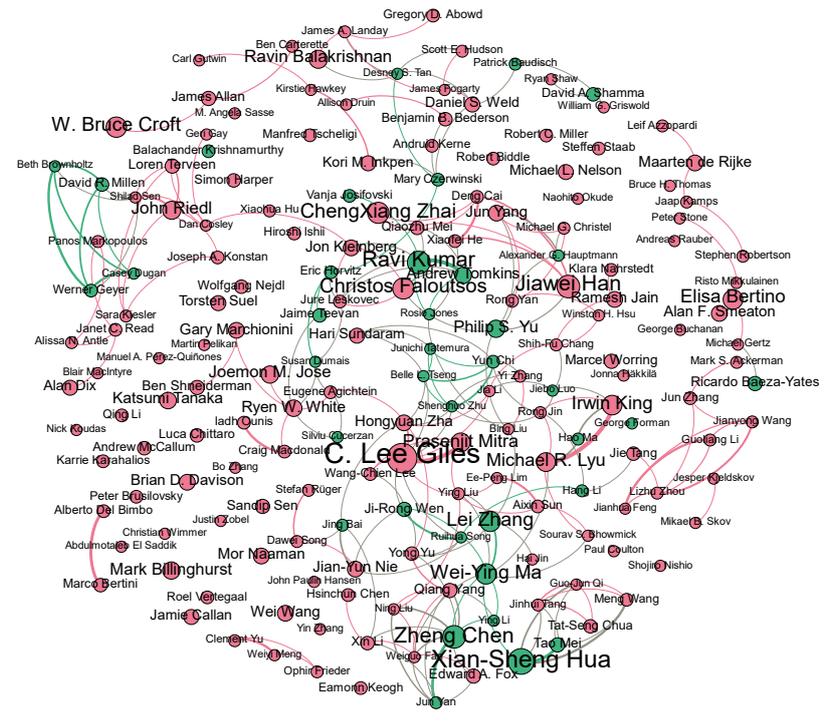

**Figure 9. Co-authorship network 2005-2009 (Top: node color by modularity; bottom: node color by authors' categories).**

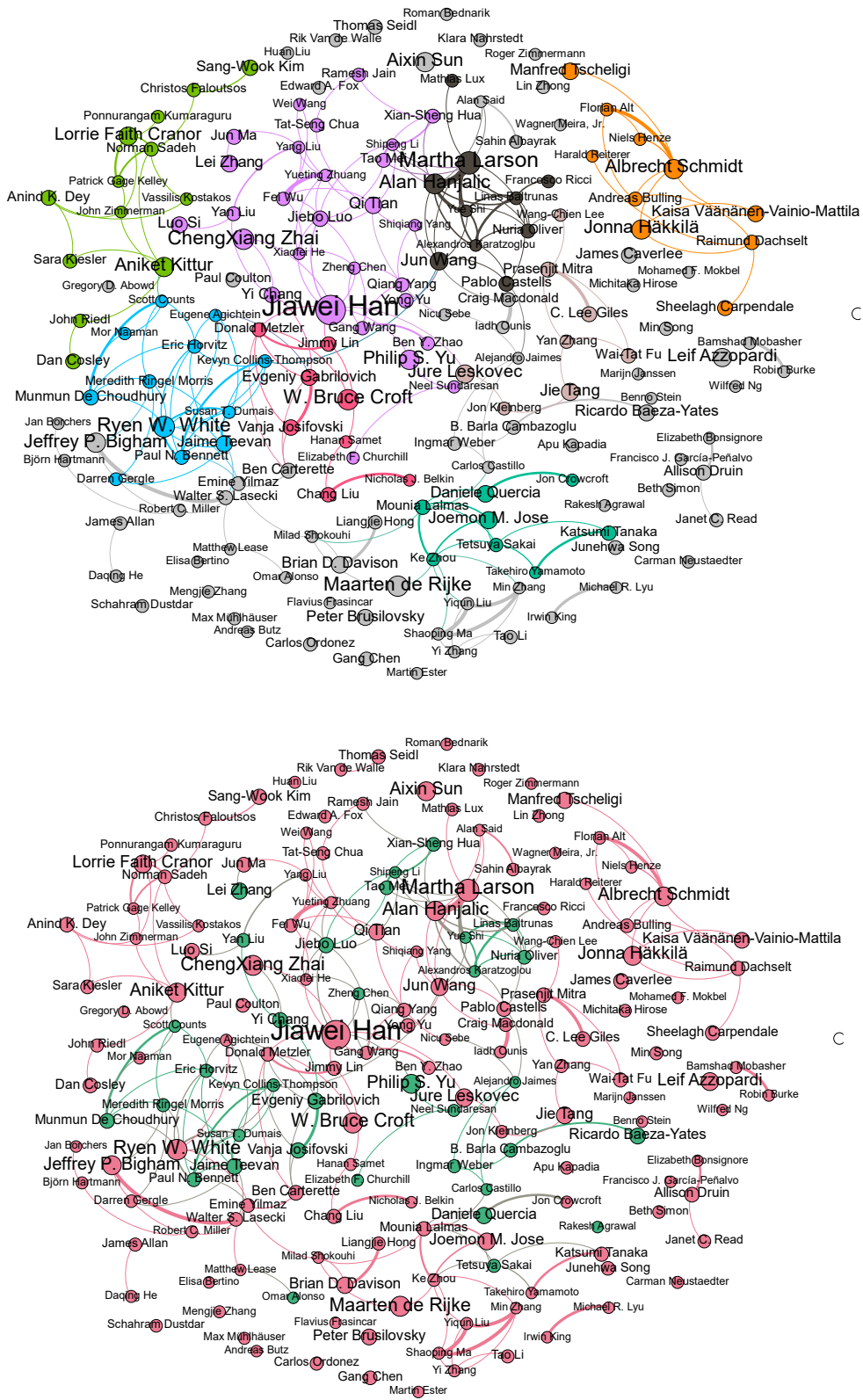

**Figure 10. Co-authorship network 2010-2014 (Top: node color by modularity; bottom: node color by authors' categories).**

Figure 11. Co-authorship network 2015-2018 (Top: node color by modularity; bottom: node color by authors' categories).

# DISCUSSION

In this paper, we explore four aspects of IR research: productivity patterns and preferred venues, the relationship between citations and downloads, changes in research topics, and changes in patterns of scientific collaboration, by analyzing and comparing publication pairs from both industrial and academic researchers in the field of information retrieval. In terms of practical implication, our findings indicate that the inclusion of authors from industry makes the research topics more oriented toward practical applications. The relationship between citations and downloads is similar for Academia and Academia-Industry Collaboration, but differs more significantly for Industry, with Academia-Industry Collaboration more likely to achieve higher download conversion rates, suggesting that collaboration can increase the impact of research. By examining how gender composition affects the citations and download of papers, we found that female scientists are relatively disadvantaged in scientific research of IR regardless of Academia, Industry, or Academia-Industry collaboration. Through topic analysis, we found that: Academic research covers diverse topics, while industry research focuses on specific platforms, data formats, and commercial applications; Collaborations between academia and industry involve both algorithmic studies and tool- or dataset-specific experiments; Also, recent academia-industry collaborative research pays increasing attention to human-centered challenges, research topics and methods, such as cyberbullying, chatbot development, and crowdsourcing analytics. Among the collaboration models, Academia-Industry Collaboration is more oriented towards large teamwork. Further mapping of the collaboration network between researchers in academia and industry reveals from the diagram that the field of IR has become richer over time in terms of themes, focus and subtopics, turning into a more diverse field of research. These conclusions help the Industry and Academia understand each other's research characteristics, thereby better promoting practical cooperation between them.

Undoubtedly, this study has certain limitations that should be addressed. Firstly, our dataset only covers information up to 2018, and considering that IR is constantly evolving, there may have been significant developments in recent years. Acquiring more up-to-date data would enhance the robustness of our conclusions and strengthen their validity. Secondly, IR review conferences have been a driving force in setting the research agenda in IR. However, since few co-authored papers on planning are included in our chosen dataset, we may not be able to get an accurate and complete understanding of the history of IR when exploring the evolution of IR and the change of research topics. Additionally, since we did not have access to full-text data, our analysis of research topics was confined to title and abstract pairs. Having complete data and more fine-grained features would have allowed for a more comprehensive exploration of the changes in research topics and team structures over time across the three types of papers. More data can also help supplement control variables in the regression model, such as the impact factor of the journal where the paper is published, the average citation number of all the authors, etc., to enhance the robustness of the regression. Furthermore, when examining scientific collaboration patterns, we opted to limit the number of nodes to focus on key contributors and extract pertinent information from the collaboration network graph. However, this approach might have resulted in some relevant information being overlooked.

In addition, ACM dataset do not contain some journal papers in the field of information retrieval. However, we initially used the MAG dataset to find areas related to the field of information retrieval, which involves a smaller content of papers, and most of the papers studying information retrieval are published as conference papers. Therefore, ACM was chosen as a more complete reflection of the history of the field of information retrieval than MAG.

Despite these limitations, our research still holds significant theoretical implications. We employed a combination of strategies, such as utilizing ACM and matching multiple datasets, to obtain the most comprehensive and accurate dataset possible, and serves as an initial step toward understanding how collaborations, research topics, and productivity evolve over time in IR community, a key interdisciplinary field in computing research. Moreover, by experimenting with multiple thresholds in the mapping of collaborative networks, we aimed to retain as much valuable information as possible. These methodological approaches have partially addressed the aforementioned limitations and contributed to an improved understanding of the patterns of academia-industry collaboration in information retrieval research.

This study provides a more comprehensive exploration of academic-industry collaboration in IR in terms of content, citations, and modes of collaboration. For researchers in the field of information retrieval, this paper reveals the impact and benefits of collaboration between academia and industry, encourages active collaboration between researchers in both fields and advances science in IR; for researchers studying the patterns of collaboration between academia and industry, this paper differs from other articles that start with industrial topics, and defines the research area in a field that is "binational" in nature - information retrieval - providing new research ideas and directions. To further enhance the study, future research should aim to acquire more recent data, access complete article texts, and employ advanced techniques for a more nuanced analysis of collaboration patterns and research topics. Because of the limited content of the data we had access to, our definition of collaboration was limited to cases where scientists co-authored papers, ignoring other possible forms of collaboration, such as graduate students doing internships in companies, professors acting as consultants to companies, etc. In future research, one could try to explore richer collaborations that might raise more interesting questions and provide some hands-on experience.

# ACKNOWLEDGMENTS

An early version of this paper was presented at iConference 2023. This paper was in part supported by the National Natural Science Foundation of China (#72104007 and #72174016) and the Fundamental Research Funds for the Central Universities, Peking University, China.

# APPENDIX

**Table A1. ACM article keywords (ranked by word frequency)**

| Keywords | | | |
|---|---|---|---|
| information retrieval | music | faceted search | graph mining |
| recommender systems | cloud computing | semantics | measurement |
| evaluation | ontologies | experimentation | semantic relatedness |
| collaborative filtering | annotation | social tagging | supervised learning |
| clustering | query reformulation | audio | sampling |
| web search | digital library | spam | document representation |
| personalization | query processing | query classification | mapreduce |
| social media | topic model | test collection | transfer learning |
| relevance feedback | social network analysis | group recommendation | web search engine |
| machine learning | big data | inverted index | query |
| information extraction | music information retrieval | community detection | personal information management |
| image retrieval | Twitter | location-based services | index |
| query expansion | implicit feedback | web | video analysis |
| social networks | retrieval | similarity measure | caching |
| ontology | web services | language model | e-commerce |
| digital libraries | similarity search | mobile computing | credibility |
| semantic web | topic modeling | neural networks | video summarization |
| data mining | image search | data integration | sensor networks |
| ranking | collaboration | federated search | query intent |
| text mining | query log analysis | video | augmented reality |
| classification | tagging | PageRank | convolutional neural networks |
| search | relevance | usability | cross-language information retrieval |
| twitter | event detection | web service | time series |
| metadata | context-awareness | content-based retrieval | e-government |
| question answering | keyword search | relation extraction | XML retrieval |
| recommendation | geographic information retrieval | novelty | interaction |
| learning to rank | information seeking | re-ranking | lifelogging |
| deep learning | hashing | enterprise search | xml |
| indexing | information visualization | performance evaluation | cross-modal retrieval |
| sentiment analysis | language models | sponsored search | log analysis |
| context | interoperability | user behavior | wiki |
| diversity | trust | events | document retrieval |
| linked data | multimedia retrieval | machine translation | online social networks |
| matrix factorization | active learning | flickr | query performance prediction |
| natural language processing | scalability | user profiling | passage retrieval |
| crowdsourcing | digital humanities | collaborative tagging | emotion |
| search engine | video annotation | personalized search | user interaction |
| recommender system | query suggestion | mobile devices | named entity recognition |
| video retrieval | recommendation system | similarity | neural network |
| text classification | digital preservation | filtering | word embeddings |
| privacy | RDF | music recommendation | correlation |
| social network | MapReduce | SVM | kernel methods |
| wikipedia | browsing | recommendation systems | human-computer interaction |
| visualization | mobile | learning | image classification |
| user modeling | social search | pseudo relevance feedback | tags |
| content-based image retrieval | web 2.0 | semantic similarity | diversification |
| optimization | knowledge management | dimensionality reduction | P2P |
| web mining | Wikipedia | knowledge base | world wide web |
| search engines | user interface | content-based filtering | open data |
| topic models | document clustering | CBIR | complex event processing |
| multimedia | information filtering | user interfaces | annotations |
| peer-to-peer | language modeling | video search | database |
| feature selection | children | text categorization | retrieval models |
| semantic search | efficiency | user profile | benchmark |
| link analysis | test collections | locality sensitive hashing | query logs |
| XML | pagerank | pseudo-relevance feedback | conversational information retrieval |
| user study | feature extraction | distributed information retrieval | spoken search system |
| user studies | interactive information retrieval | blog | fairness information retrieval |
| opinion mining | expert finding | unsupervised learning | accountability information retrieval |
| summarization | search behavior | entity linking | transparency information retrieval |
| exploratory search | eye tracking | content analysis | ethnics information retrieval |
| image annotation | performance | adaptation | explainability information retrieval |
| semi-supervised learning | education | random walk | responsible information retrieval |
| folksonomy | navigation | query formulation | |

**Table A2. Research topics over time**

| | Academia | Industry | Academia-Industry Collaboration |
|---|---|---|---|
| 2000 | intelligent libraries | video classroom | learning algorithms |
|      | library classes<br>library technologies<br>novel browser<br>internet classrooms                      | video watermarking<br>video performance<br>video technical<br>video recording                 | analysis hashing<br>proliferation internet<br>classifies algorithms<br>learning algorithm          |
| 2001 | mouse popular<br>mouse 3d<br>popular multiplayer<br>3d computing<br>powerful 3d                      | advanced algorithms<br>computationally feasible<br>researchers improve<br>modeling useful<br>interface powerful | software engineers<br>designer needs<br>designing ontology<br>xml software<br>documentation engineers |
| 2002 | libraries tutorial<br>databases attractive<br>library technology<br>efficient indexing<br>databases tutoring | designing web<br>auctions improving<br>expanding rehearsal<br>bioinformatics emerging<br>search engines | offering algorithms<br>tackling algorithms<br>semantic web<br>algorithms software<br>web query |
| 2003 | optimization cancer<br>audition algorithms<br>partitioning algorithms<br>algorithms comparison<br>algorithms haplotype | music photo2video<br>retrieves songs<br>music concert<br>music database<br>new songs | simplifying web<br>internet experiments<br>popular web<br>online semantic<br>new spam |
| 2004 | algorithm updating<br>novel algorithms<br>algorithms lessons<br>developing algorithms<br>algorithms learning | toolkit debugging<br>browser optimizations<br>search algorithms<br>web verification<br>algorithms methodology | simplifying web<br>internet experiments<br>popular web<br>online semantic<br>new spam |
| 2005 | valuable indexing<br>algorithms improving<br>efficient ontology<br>interesting research<br>important crosscutting | algorithms methodology<br>magic instructional<br>data webgazeanalyzer<br>webgazeanalyzer brings<br>laser pointer | holistic algorithms<br>novel internetworking<br>smart phones<br>algorithms scalability<br>wireless broadband |
| 2006 | valuable indexing<br>algorithms improving<br>efficient ontology<br>interesting research<br>important crosscutting | research challenging<br>executed twice<br>algorithms counter<br>bioinformatics motivation<br>steep learning | challenge traffic<br>traffic engineering<br>algorithms damping<br>algorithms research<br>engineering algorithms |
| 2007 | servers cheating<br>privacy vulnerabilities<br>online personalization<br>leakage internet<br>online anonymity | huge database<br>hashing billions<br>largest commerce<br>amazon highly<br>huge collections | major browsers<br>algorithms large<br>algorithms widely<br>extensive programming<br>fastdash developer |

| Year | | | |
|---|---|---|---|
| 2008 | wikipedia huge<br>browsers popular<br>huge databases<br>favourite websites<br>winning podcasts | study robots<br>survey robots<br>querybuilder query<br>querylogs bundling<br>botnet detection | algorithm recommender<br>study novel<br>research recent<br>tagging podcasts<br>researchers paper |
| 2009 | rfid popular<br>patents important<br>rfid algorithms<br>patents essential<br>tagging expert | browsers actively<br>growth online<br>online advertising<br>online surveys<br>pushing browsers | internet researchers<br>importance researchers<br>benchmarking browsers<br>research researchers<br>data researchers |
| 2010 | new algorithms<br>hashtag innovation<br>evolving wiki<br>new apps<br>discovery bioinformatics | research revolutionize<br>expensive simulators<br>database huge<br>budget challenging<br>consumption skyrocketed | wikiprojects increased<br>pagerank algorithm<br>playlists photoselect<br>brainstorming stylus<br>retagging online |
| 2011 | design tutorial<br>bioinformaticians designing<br>designing sparql<br>clustering bioinformatics<br>tutorials technologies | driver safety<br>automobiles tutorial<br>refocus driving<br>driver infotainment<br>opportunistic driver | attractive websites<br>understanding internet<br>moderating online<br>good websites<br>modeling internet |
| 2012 | comics techcommix<br>cancer increased<br>cyberinfrastructure scientist<br>algorithms physicists<br>scientists seeking | improving tweet<br>threefold allows<br>reducing aliasing<br>algorithm reducing<br>strategies threefold | avatar conferencing<br>rearranging videos<br>video hashing<br>video tutorials<br>media tutorials |
| 2013 | detectives solving<br>algorithms greedy<br>played detectives<br>crime notepad<br>detecting cyberbullying | latest poll<br>proposed algorithm<br>tweeted wedding<br>motivation graphbuilder<br>research innovations | project openstreetmap<br>tutorial overview<br>openstreetmap editors<br>pattern openstreetmap<br>modeling tutorial |
| 2014 | discriminating online<br>misinformation<br>crowdturfing<br>instagram traffic<br>kickstarter interview<br>internet challenging | improving online<br>improvements online<br>google overtaking<br>web analytics<br>wikipedia benefitted | pagerank algorithm<br>web searchers<br>search engines<br>search engine<br>online surveys |
| 2015 | apps revolutionize<br>apps changing<br>simplifying mobile<br>investigating smartphone | improve genetic<br>experiments expensive<br>expensive experiments<br>novel algorithms | youtube tutorials<br>movielens netflix<br>netflix datasets<br>netflix dataset |

|  | developing smartwatch | quantitative genetic | youtube flickr |
|---|---|---|---|
| 2016 | learning analytics<br>analytics learning<br>agile analytics<br>learning smartwatches<br>learning evolution | google facebook<br>facebook microsoft<br>facebook conducting<br>traffic staggering<br>economics rigorously | chatbots developers<br>needed mobile<br>designing android<br>prototype chatbot<br>apps study |
| 2017 | videos rebooting<br>videoconferencing application<br>video tutorials<br>video study<br>videoconferencing | google facebook<br>facebook microsoft<br>facebook conducting<br>traffic staggering<br>economics rigorously | online bullying<br>facebook misleading<br>bullying twitter<br>online distractions<br>reducingcontroversy homepage |
| 2018 | videos rebooting<br>videoconferencing application<br>video tutorials<br>video study<br>videoconferencing | important research<br>seismic interpreters<br>cryptography needed<br>noisy training<br>research challenges | crowdsourcing analytics<br>reuse networking<br>cache management<br>cache reuse<br>web transformational |

Table A3. Perplexity indexes with different number of topics.

| # topic | Perplexity index | # topic | Perplexity index |
|---|---|---|---|
| 1 | 129.113 | 8 | 111.642 |
| 2 | 122.302 | 9 | 111.476 |
| 3 | 118.342 | 10 | 111.862 |
| 4 | 115.826 | 11 | 111.581 |
| 5 | 113.819 | 12 | 111.254 |
| 6 | 112.655 | 13 | 111.253 |
| 7 | 112.118 | 14 | 110.369 |

Table A4. Node filtering results for each time cohort.

| Time Period | Minimum degree of nodes | Number of remaining nodes | Number of clusters |
|---|---|---|---|
| 2000-2004 | 4 | 186 | 99 |
| 2005-2009 | 8 | 184 | 69 |
| 2010-2014 | 12 | 170 | 53 |
| 2015-2018 | 9 | 167 | 77 |